%% file: manuscript_arxiv.tex
\documentclass[journal]{IEEEtran}
\IEEEoverridecommandlockouts

\input{addons/preamble}

\renewcommand{\baselinestretch}{1}

\newcommand{\ten}[1]{\boldsymbol{\mathcal #1}}
\newcommand{\nmode}[2]{\big[\boldsymbol{{\mathcal{#1}}}\big]_{\left(#2\right)}}

\pdfoutput=1

\begin{document}

\bstctlcite{IEEE:BSTcontrol}

\begin{acronym}[MPC] 
	
	\acro{6G}{sixth generation} 
	
	\acro{AR}{auto-regressive}
	
	\acro{AoA}{angle of arrival}
	
	\acro{AoD}{angle of departure}
	
	\acro{AWGN}{additive white Gaussian noise}
	
	\acro{B5G}{beyond fifth generation}
	
	\acro{BER}{bit error ratio}
	
	\acro{BS}{base station}
	
	\acro{BPSK}{binary phase-shift keying}
	
	\acro{CRLB}{Cramér-Rao lower bound}
	
	\acro{gNB}{next generation node B}
	
	\acro{IRS}{intelligent reflecting surface}
	
	\acro{ISAC}{integrated sensing and communications}
	
	\acro{MIMO}{multiple input multiple output}
	
	\acro{CSI}{channel state information}
	
	\acro{DFT}{discrete Fourier transform}
	
	\acro{GLRT}{generalized likelihood ratio test}
	
	\acro{ISAC}{integrated sensing and communications}
	
	\acro{LOS}{line of sight}
	
	\acro{MMSE}{minimum mean squared error}
	
	\acro{NMSE}{normalized mean squared error}
	
	\acro{NLOS}{non-line of sight}
	
	\acro{OFDM}{orthogonal frequency division multiplexing}
	
	\acro{RCS}{radar cross section}
	
	\acro{RMSE}{root mean squared error}
	
	\acro{SNR}{signal-to-noise ratio}
	
	\acro{SVD}{singular value decomposition}
	
	\acro{SE}{spectral efficiency}
	
	\acro{UE}{user equipment}
	
	\acro{ULA}{uniform linear array}
	
	\acro{URA}{uniform rectangular array}
	
	
	\acro{HOSVD}{higher-order singular value decomposition}
	
\end{acronym}

\title{Low-Complexity Tensor-Based Monostatic Sensing for IRS-Assisted Communication Systems\\
{\footnotesize}}

\author{Kenneth B. A. Benício, Bruno Sokal, André L. F. de Almeida, Fazal-E-Asim, Behrooz Makki, and Gábor Fodor}

\maketitle

\begin{abstract}
     This paper proposes a tensor-based parameter estimation algorithm for sensing in an intelligent reflecting surface-assisted system. We present a higher-order singular value decomposition-based solution that exploits the tensor structure of the received echo signal to jointly estimate the target's delay, Doppler, and angular information. Our tensor-based solution can estimate the parameters individually at low complexity, benefiting from parallel computation. Complexity analysis is carried out in comparison with a baseline scheme that does not exploit the intrinsic multilinear structure of the sensed signal. Simulation results show that our proposed tensor-based method can achieve the same performance as the reference method while drastically reducing the computational complexity.
\end{abstract}


\acresetall

\section{Introduction}
    
    In recent years, the \acp{IRS} implementation has been vastly researched due to its promising capabilities of improving the overall system throughput, which has the potential to enable the desirable standards of the \ac{6G} of mobile communications \cite{zheng2022survey, pan2022overview, astrom2024ris}. The \ac{IRS} is a two-dimensional planar object composed of a high number of passive reflecting elements capable of individually changing the phase shifts of impinging electromagnetic waves to smartly maximize the \ac{SNR} at some intended receiver \cite{wu2021intelligent}.

    In this context, radar sensing and communication systems are advancing towards elevated frequency bands, increasing the number of antennas that allow miniaturization (smaller size arrays) and signal processing. This convergence presents a compelling prospect for integrating sensing functionalities into wireless infrastructures in future networks. Although this functionality integration has been investigated under different names over the past decades, it has only caught the attention of academia and industry in the past couple of years. This radar sensing and communication integration is commonly known as \ac{ISAC} \cite{liu2022integrated, chepuri2023integrated}. Here, by ISAC, the objective is to share the spectrum more efficiently and/or reuse the existing wireless network infrastructure for sensing. Based on this concept, the \ac{IRS} arises as an interesting tool for \ac{ISAC} systems due to its capability of smartly reshaping the propagation environment to enhance system metrics and also illuminating \ac{NLOS} areas. In this work, we utilize the \acp{IRS} to enhance the sensing aspects of the network while employing a communications waveform to sense the channel characteristics. Also, due to the passive essence of the \ac{IRS}, any sensing procedure must be implemented at the end nodes of the network by using the received echo signal reflected by the \ac{IRS}. Several works have already addressed this problem, as the ones mentioned in \cite{zhang2022metaradar, rihan2022spatial, song2023intelligent, shao2024target, kemal2024ris}

    In the work of \cite{zhang2022metaradar}, the \ac{IRS} is employed as a means to improve the detection performance of a multi-target \ac{MIMO} radar system while jointly optimizing the radar waveform and the \ac{IRS} phase-shift. The authors in \cite{rihan2022spatial} consider that the \ac{IRS} is deployed to increase the spatial diversity of a radar system by creating an additional virtual \ac{LOS} path between the radar and the intended target. In \cite{song2023intelligent}, the authors investigate an \ac{IRS}-assisted \ac{NLOS} sensing scenario and propose the active and passive beamforming design that minimizes the \ac{CRLB}. The work \cite{shao2024target} proposes a secure sensing scenario with a target-mounted \ac{IRS} that aims to enhance the sensing performance of the overall system while preventing detection of the target by potential eavesdroppers. The authors in \cite{kemal2024ris} propose a parameter estimation method based on the design of a \ac{GLRT} detector, which involves the computation of a four-dimensional parameter peak search of delay-Doppler-azimuth-elevation. To simplify the computation, the authors employ a repetitive \ac{IRS} phase-shift profile that breaks the four-dimensional search into a delay-Doppler and azimuth-elevation search problems. 
    
    In this paper, we develop low-complexity monostatic sensing-based methods in IRS-assisted networks. Notably, we propose a reinterpretation of the monostatic sensing \ac{IRS}-assisted scenario proposed in \cite{kemal2024ris}. The solution of this paper resorts to tensor methods to recast the received echo signal at the \ac{BS} as a third-order rank-one tensor from which we solve the parameter estimation problem by employing the well-known \ac{HOSVD} algorithm. We perform a complexity analysis of the proposed scheme and study the effect of different parameters on the performance of IRS-assisted ISAC networks. Our simulation results show that the proposed tensor-based sensing solution performs similarly to the benchmark solution of \cite{kemal2024ris} in terms of \ac{RMSE} while substantially reducing the computational complexity to solve the parameter estimation problem.
    
    \textit{Notation}: Scalars, vectors, matrices, and tensors are represented as $a, \boldsymbol{a}, \boldsymbol{A}$, and $\ten{A}$. Also, $\boldsymbol{A}^{*}$, $\boldsymbol{A}^{\text{T}}$, $\boldsymbol{A}^{\text{H}}$, and $\boldsymbol{A}^{\dagger}$ stand for the conjugate, transpose, Hermitian, and pseudo-inverse, of a matrix $\boldsymbol{A}$. The $j$th column of $\boldsymbol{A} \in \mathbb{C}^{I \times J}$ is denoted by $\boldsymbol{A}_{.j} \in \mathbb{C}^{I \times 1}$. The operator D$(\cdot)$ converts a vector into a diagonal matrix, $\text{D}_j(\boldsymbol{B})$ forms a diagonal matrix $R \times R$ out of the $j$th row of $\boldsymbol{B} \in \mathbb{C}^{J \times R}$. Also, $\boldsymbol{I}_{N}$ denotes an identity matrix of size $N \times N$. The symbols $\otimes$, $\diamond$, and $\odot$ indicate the Kronecker, Khatri-Rao, and Hadamard products.
    
\section{System Model}
    \begin{figure}
        \centering
        \begin{tikzpicture}[scale=1, every node/.style={scale=0.975}]
            \begin{scope}[
                box1/.style={draw=black, thick, rectangle,rounded corners, minimum height=0.5cm, minimum width=0.5cm}]
                \node (IRS 1) at (-2,0.75) {$\text{IRS}$};
                \draw[black,dashed,fill=red!30] (-3.5,-2.5) rectangle (-.5,.5);
                \node[box1, fill=green!30] (c1) at (-3.,0) {};
                \node[box1, fill=green!30, right=.125cm of c1] (c2) {};
                \node[box1, fill=green!30, right=.125cm of c2] (c3) {};
                \node[box1, fill=green!30, right=.125cm of c3] (c4) {};
                \node[box1, fill=green!30, below=.125cm of c4] (c5) {};
                \node[box1, fill=green!30, left=.125cm of c5] (c6) {};
                \node[box1, fill=green!30, left=.125cm of c6] (c7) {};
                \node[box1, fill=green!30, left=.125cm of c7] (c8) {};
                \node[box1, fill=green!30, below=.125cm of c8] (c9) {};
                \node[box1, fill=green!30, right=.125cm of c9] (c10) {};
                \node[box1, fill=green!30, right=.125cm of c10] (c11) {};
                \node[box1, fill=green!30, right=.125cm of c11] (c12) {};
                \node[box1, fill=green!30, below=.125cm of c12] (c13) {};
                \node[box1, fill=green!30, left=.125cm of c13] (c14) {};
                \node[box1, fill=green!30, left=.125cm of c14] (c15) {};
                \node[box1, fill=green!30, left=.125cm of c15] (c15) {};
                \draw[black,fill=blue!30] (-7+1.5,-3.5) -- (-6.5+1.5,-3.5) node[above]{$\text{BS}$} -- (-6+1.5,-3.5) -- (-6.5+1.5,-2.4) -- cycle;
                \node[circle,draw=black,fill=blue!30,minimum size=12pt] (T1) at (2.70-2,-2-1) {};
                \node[above] (Target) at (2.70-2,-1.5-1.26) {$\text{Target}$};
                \draw[line width=0.25mm,black] (-7+1.5,-2.4) -- (-6+1.5,-2.4) node[midway,above]{$\cdots$};
                \draw[line width=0.25mm,black] (-7+1.5,-2.4) -- (-7+1.5,-2.35);
                \draw[line width=0.25mm,red] (-6+1.5,-2.4) -- (-6+1.5,-2.35);
                \draw[line width=0.25mm,black] (-7.20+1.5,-2.25) -- (-7+1.5,-2.35) -- (-6.80+1.5,-2.25) -- cycle;
                \draw[line width=0.25mm,red] (-6.20+1.5,-2.25) -- (-6+1.5,-2.35) -- (-5.80+1.5,-2.25) -- cycle;
                \draw[line width=0.75mm,black,->] (-6.5+1.5,-1.65) -- (-3.75,0) node[midway, above, rotate=+50]{};
                \draw[line width=0.75mm,red,<-] (-6.5+1.5,-1.95) -- (-3.75,-0.30) node[midway, above, rotate=+50]{};
                \draw[line width=0.75mm,black,->] (-0.25,0) -- (2-1.15,-1.65);
                \draw[line width=0.75mm,red,<-] (-0.25,-0.30) -- (2-1.15,-1.95);
                \draw[line width=0.75mm,dashed,black!75,->] (-6+1.5,-3) -- (-2.25,-3) node[midway, above, rotate=+50]{};
                \draw (-1.95,-3) pic[rotate = 0] {cross=7pt};
            \end{scope}
        \end{tikzpicture}
        \caption{IRS-assisted SISO Monostatic Sensing.}
        \label{fig:system_model}
    \end{figure}
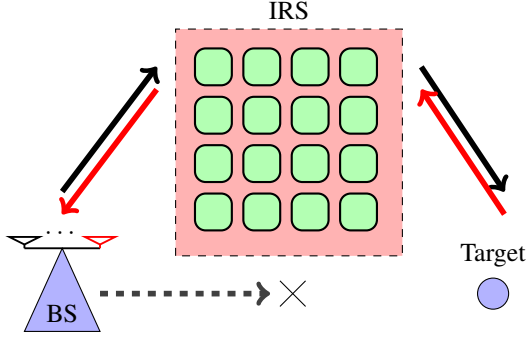
    Our system model revisits the scenario proposed in \cite{kemal2024ris}. Specifically, we consider a downlink monostatic \footnote{In a monostatic sensing scenario, the self-interference is a natural issue present \cite{liu2022integrated}. In this work, the self-interference is out of scope as for the baseline state-of-the-art competitor \cite{kemal2024ris}, since we aim to provide a lower complexity solution for the problem of \cite{kemal2024ris}.} \ac{IRS}-aided scenario under \ac{LOS} blockage with a dual-functional \ac{BS} with a single antenna dedicated to transmit an \ac{OFDM} waveform and receives the echo signal backscattered by the target through the \ac{IRS}. We consider a passive \ac{IRS} equipped with $N$ reflecting elements. Furthermore, the \ac{IRS} localization and orientation are assumed to be known, which is an acceptable assumption given that, in practice, the IRS has a fixed position. The \ac{BS} transmits an \ac{OFDM} signal with $N_{c}$ subcarriers and $M$ symbols, with $x_{n_{c},m}$ denoting the complex data/pilot on the $n_{c}$th subcarrier of the $m$th symbol. Assuming the existence of a single target, the received \ac{OFDM} echo signal at the \ac{BS}, switching to frequency-domain, is expressed as \cite{kemal2024ris}
    \begin{align}
        \begin{split} y_{n_{c},m} = \alpha \underbrace{\boldsymbol{a}^{\text{T}}(\boldsymbol{\phi}) \text{D}(\boldsymbol{w}_{m}) \boldsymbol{p}(\boldsymbol{\theta})}_{\text{Target-\ac{IRS}-\ac{BS} path}} \underbrace{\boldsymbol{p}^{\text{T}}(\boldsymbol{\theta}) \text{D}(\boldsymbol{w}_{m}) \boldsymbol{a}(\boldsymbol{\phi})}_{\text{\ac{BS}-\ac{IRS}-Target path}}   \\ \times \left[\boldsymbol{c}(\tau)\right]_{n_{c}} \left[\boldsymbol{d}(\nu)\right]_{m} x_{n_{c},m} + z_{n_{c},m}, \end{split} \label{eq:received_echo_1}
    \end{align}
    \noindent where $\boldsymbol{\phi} = \left[\phi_{\text{az}}, \phi_{\text{el}}\right]^{\text{T}}$ is the \ac{AoA} from the \ac{BS} to the \ac{IRS}, $\boldsymbol{\theta} = \left[\theta_{\text{az}}, \theta_{\text{el}}\right]^{\text{T}}$ is the \ac{AoD} from the \ac{IRS} to the target, and $\boldsymbol{a}(\boldsymbol{\phi}) \in \mathbb{C}^{N \times 1}$ and $\boldsymbol{p}(\boldsymbol{\theta}) \in \mathbb{C}^{N \times 1}$ are the two-dimensional steering vectors expressed as \cite{Asim_2021}
    \begin{align}
        \boldsymbol{a}(\boldsymbol{\phi}) &= \boldsymbol{a}(\phi_{\text{az}}, \phi_{\text{el}}) \otimes \boldsymbol{a}(\phi_{\text{el}}) \in \mathbb{C}^{N \times 1}, \\
        \boldsymbol{p}(\boldsymbol{\theta}) &= \boldsymbol{p}(\theta_{\text{az}}, \theta_{\text{el}}) \otimes \boldsymbol{p}(\theta_{\text{el}}) \in \mathbb{C}^{N \times 1},
    \end{align}
    \noindent with $\phi_{\text{az}}$ and $\phi_{\text{el}}$ being the azimuth and elevation \ac{AoA}, and  $\theta_{\text{az}}$ and $\theta_{\text{el}}$ being the azimuth and elevation \ac{AoD}, respectively. Also, $\boldsymbol{w}_{m} \in \mathbb{C}^{N \times 1}$ is the \ac{IRS} phase-shift profile for the $m$th \ac{OFDM} symbol defined as $\boldsymbol{w}_{m} = \left[e^{j \zeta_{1,m}} , \cdots, e^{j \zeta_{N,m}} \right]^{\text{T}} \in \mathbb{C}^{N \times 1}$, where $\zeta_{n,m}$ is the phase-shift of the $n$th IRS element at the $m$th \ac{OFDM} symbol. The frequency-domain steering vector as a function of the delay $\tau$ is given by $\boldsymbol{c}(\tau) \in \mathbb{C}^{N_{c} \times 1}$ and the element linked to the $n$th \ac{OFDM} subcarrier is $\left[\boldsymbol{c}(\tau)\right]_{n_{c}} \in \mathbb{C}$. Also, $\boldsymbol{d}(\nu) \in \mathbb{C}^{M \times 1}$ is the time-domain steering vector as function of the Doppler $\nu$ and the element linked to the $m$th \ac{OFDM} symbol is $\left[\boldsymbol{d}(\nu)\right]_{m} \in \mathbb{C}$, and $z_{n_{c},m} \in \mathbb{C}$ is the \ac{AWGN} component at the $n$th subcarrier of the $m$th symbol. Finally, the magnitude of the complex gain, $\alpha \in \mathbb{C}$, is given by
    \begin{align}
        |\alpha| = \sqrt{\frac{P_{t} G^{2}_{1} G^{2}_{2} F^{2}_{1}(\phi_{\text{el}}) F^{2}_{2}(\theta_{\text{el}}) d^{2}_{x} d^{2}_{y} \lambda^{2} \sigma_{\text{RCS}}}{(4\pi)^{5} d^{4}_{1}d^{4}_{2}}},
    \end{align}
    \noindent where $P_{t}$ is the transmit power, $G_{1}$ is the \ac{BS} transmit antenna gain, $G_{2}$ is the \ac{BS} receive antenna gain, $F^{2}_{1}(\phi_{\text{el}})$ is the normalized \ac{IRS} power radiation pattern at the \ac{AoA} $\phi_{\text{el}}$, $F^{2}_{2}(\theta_{\text{el}})$ is the normalized \ac{IRS} power radiation pattern at the \ac{AoD} $\theta_{\text{el}}$, $d_{x}$ and $d_{y}$ denotes the \ac{IRS} spacing along the horizontal and vertical domains, respectively, $\lambda$ is the carrier wavelength, $\sigma_{\text{RCS}}$ is the target \ac{RCS}, and $d_{1}$ and $d_{2}$ denotes the distances \ac{BS}-\ac{IRS} and \ac{IRS}-target, respectively. To compact Equation (\ref{eq:received_echo_1}), we can define $\boldsymbol{b}(\boldsymbol{\phi}, \boldsymbol{\theta}) = \boldsymbol{a}(\boldsymbol{\phi}) \odot \boldsymbol{p}(\boldsymbol{\theta})$ and recast the expression of the echo signal as
    \begin{align}
        y_{n_{c},m} &= \alpha (\boldsymbol{b}^{\text{T}}(\boldsymbol{\phi},\boldsymbol{\theta}) \boldsymbol{w}_{m})^{2}  \left[\boldsymbol{c}(\tau)\right]_{n_{c}}  \left[\boldsymbol{d}(\nu)\right]_{m} x_{n_{c},m} + z_{n_{c},m} \label{eq:received_echo_xxx}.
    \end{align}
   
    To stack over $N_{c}$ subcarriers and $M$ symbols, we define $\boldsymbol{W} \in \mathbb{C}^{N \times M}$ as a matrix that collects the \ac{IRS} phase-shift patterns during $M$ \ac{OFDM} symbols. We can recast the received signal in Equation (\ref{eq:received_echo_xxx}) as \footnote{The design of the phase-shift matrix is discussed in Section III A.}
    \begin{align}
        \begin{split} \boldsymbol{Y} = \boldsymbol{X} \odot \alpha \boldsymbol{c}(\tau)\left(\boldsymbol{d}^{\text{T}}(\nu) \odot \boldsymbol{b}^{\text{T}}(\boldsymbol{\phi},\boldsymbol{\theta}) \boldsymbol{W} \odot \boldsymbol{b}^{\text{T}}(\boldsymbol{\phi},\boldsymbol{\theta}) \boldsymbol{W} \right) \\ + \boldsymbol{Z} \in \mathbb{C}^{N_{c} \times M},\end{split}   
    \end{align}
    \noindent where $\boldsymbol{c}(\tau) \in \mathbb{C}^{N_{c} \times 1}$ is the complete frequency-domain steering vector as a function of the delay $\tau$, $\boldsymbol{d}(\nu) \in \mathbb{C}^{M \times 1}$ is the time-domain steering vector as a function of the Doppler $\nu$, $\boldsymbol{X} \in \mathbb{C}^{N_{c} \times M}$ is the transmitted symbols matrix by the \ac{BS} where $\boldsymbol{X}_{n_{c},m} = x_{n_{c},m}$, and $\boldsymbol{Z} \in \mathbb{C}^{N_{c} \times M} \sim \mathcal{CN}(\boldsymbol{0}, \sigma \boldsymbol{I})$ is the \ac{AWGN} matrix. Also, by considering that the entries of symbols matrix $\boldsymbol{X}$ are composed of unitary elements and that the knowledge of the angular information of the channel between the \ac{BS} and \ac{IRS}, $\boldsymbol{\phi}$, is known at the receiver, then it is possible to simplify the formulation of the received echo signal at the \ac{BS} with the following expression \cite{kemal2024ris}
    \begin{align}
        \boldsymbol{Y} \hspace{-0.1cm}=\hspace{-0.05cm} \alpha \boldsymbol{c}(\tau) \hspace{-0.05cm} \left(\boldsymbol{d}^{\text{T}}(\nu) \hspace{-0.05cm}\odot\hspace{-0.05cm} \boldsymbol{b}^{\text{T}}(\boldsymbol{\theta}) \boldsymbol{W} \hspace{-0.05cm}\odot\hspace{-0.05cm} \boldsymbol{b}^{\text{T}}(\boldsymbol{\theta}) \boldsymbol{W} \right) \hspace{-0.05cm}+\hspace{-0.05cm} \boldsymbol{Z} \hspace{-0.05cm}\in\hspace{-0.05cm} \mathbb{C}^{N_{c} \hspace{-0.05cm}\times\hspace{-0.05cm} M}. \label{eq:received_echo_2}
    \end{align}
\section{Proposed Tensor-Based Sensing Receiver}
    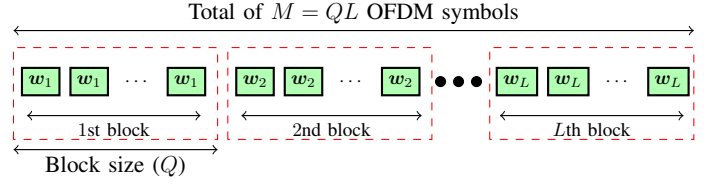
\begin{figure}[t!]
        \centering
        \begin{tikzpicture}[scale=0.90, every node/.style={scale=0.725}]
            \begin{scope}[
                box1/.style={draw=black, thick, rectangle, minimum height=0.5cm, minimum width=0.5cm}]
                \draw[black,dashed,red,fill=white] (-4.65,-.85) rectangle (-2+0.35,.5);
                \node[box1, fill=green!30] (c1) at (-4.25,0) {$\boldsymbol{w}_{1}$};
                \node[box1, fill=green!30, right=.125cm of c1] (c2) {$\boldsymbol{w}_{1}$};
                \node[right=.125cm of c2] (c3) {$\cdots$};
                \node[box1, fill=green!30, right=.125cm of c3] (c4) {$\boldsymbol{w}_{1}$};
                \draw[black,dashed,red,fill=white] (-1.50,-.85) rectangle (1.15+0.35,.5);
                \node[box1, fill=green!30] (c1) at (-1.10,0) {$\boldsymbol{w}_{2}$};
                \node[box1, fill=green!30, right=.125cm of c1] (c2) {$\boldsymbol{w}_{2}$};
                \node[right=.125cm of c2] (c3) {$\cdots$};
                \node[box1, fill=green!30, right=.125cm of c3] (c4) {$\boldsymbol{w}_{2}$};
                \node[right=.20 of c4,circle,draw=black,fill=black,minimum size=1pt,scale=0.55] (T1) {};
                \node[right=.1 of T1,circle,draw=black,fill=black,minimum size=1pt,scale=0.55] (T2) {};
                \node[right=.1 of T2,circle,draw=black,fill=black,minimum size=1pt,scale=0.55] (T3) {};
                \draw[black,dashed,red,fill=white] (2.35,-.85) rectangle (5.35,.5);
                \node[box1, fill=green!30] (c1) at (2.75,0) {$\boldsymbol{w}_{L}$};
                \node[box1, fill=green!30, right=.125cm of c1] (c2) {$\boldsymbol{w}_{L}$};
                \node[right=.125cm of c2] (c3) {$\cdots$};
                \node[box1, fill=green!30, right=.125cm of c3] (c4) {$\boldsymbol{w}_{L}$};
                \draw[line width=0.15mm,black,<->] (-4.65,-1) -- (-1.65,-1) node[midway, below, rotate=+0]{\large Block size ($Q$)};
                \draw[line width=0.15mm,black,<->] (-4.65,0.75) -- (5.35,0.75) node[midway, above, rotate=+0]{\large Total of $M= QL$ OFDM symbols};
                \draw[line width=0.15mm,black,<->] (-4.45,-0.5) -- (-2.25+0.35,-0.5) node[midway, below, rotate=+0]{$1$st block};
                \draw[line width=0.15mm,black,<->] (-1.3,-0.5) -- (1+0.35,-0.5) node[midway, below, rotate=+0]{$2$nd block};
                \draw[line width=0.15mm,black,<->] (2.5,-0.5) -- (4.85+0.35,-0.5) node[midway, below, rotate=+0]{$L$th block};
            \end{scope}
        \end{tikzpicture}
        \caption{Time-domain protocol of the proposed IRS design.}
        \label{fig:transmission_protocol}
    \end{figure}
\begin{figure}[!t]
        \centering
        \includegraphics[width=0.95\columnwidth]{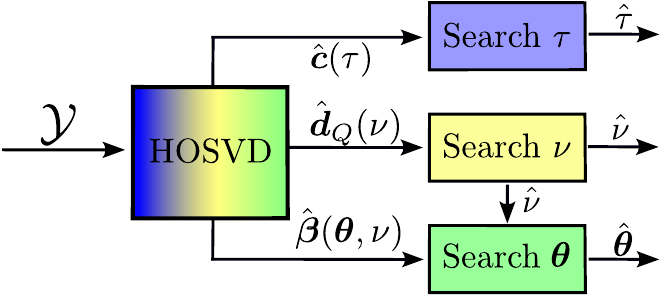}
        \caption{Block diagram of the proposed parameter estimation}
        \label{fig:03}
    \end{figure}

    This section presents our proposed tensor-based method for parameter estimation and investigates its implementation complexity. Different from \cite{kemal2024ris}, our proposed solution recast the received echo signal from the BS-IRS-target-IRS-BS link as a third-order rank-one tensor to solve the parameter estimation problem based on the state-of-the-art \ac{HOSVD}, as shown in Algorithm \ref{alg:01}. The main idea is to exploit the intrinsic multilinear structure of the received signal to estimate the best rank-one approximation of the tensor signal so that in the sequence, we can solve three individual peak search problems from which we obtain the estimates of $\tau$, $\nu$, and $\boldsymbol{\theta}$ as shown in Fig. \ref{fig:03}. Due to the tensor approach, we harvest considerable complexity gain as shown in Table \ref{tab:complexity}.
    
    \subsection{Formulation of the Proposed Tensor Model}
       From Equation (\ref{eq:received_echo_xxx}), we observe that the \ac{IRS} pattern is varying with each \ac{OFDM} symbol period, which results in a coupling between the Doppler $\nu$ and the angles $(\phi, \theta)$ as showcased in Equation (\ref{eq:received_echo_2}). To solve this, we employ a repetition-based \ac{IRS} phase-shift pattern to decouple the angular and Doppler parameter estimates, as proposed by \cite{kemal2024ris}. The idea consists of repeating the same phase-shift pattern during a block of $Q$ \ac{OFDM} symbols while varying this pattern across $L$ blocks, yielding a total time window of $M=QL$ OFDM symbols. This protocol is illustrated in Fig. \ref{fig:transmission_protocol}. As a result, the IRS phase shift matrix can be expressed as
        \begin{align}
            \boldsymbol{W} = \left[\boldsymbol{w}_{1} \cdots \boldsymbol{w}_{L}\right] \otimes \boldsymbol{1}^{\text{T}}_{Q} \in \mathbb{C}^{N \times Q L}. \label{eq:IRS_profile_2}
        \end{align}
        Since the angular parameters are not dependent on the time domain, in contrast to the Doppler parameter, such a repetition protocol allows decoupling the estimate of the pure Doppler-related vector from the angle-Doppler-related vector, as will be shown in this section. To exploit this new design for the \ac{IRS} phase-shift matrix in Equation (\ref{eq:received_echo_2}), let us define 
        \begin{align}
            \boldsymbol{g}^{\text{T}}(\boldsymbol{\theta}) = \boldsymbol{b}^{\text{T}}(\boldsymbol{\theta}) \boldsymbol{W} \odot \boldsymbol{b}^{\text{T}}(\boldsymbol{\theta}) \boldsymbol{W} \in \mathbb{C}^{1 \times Q L},
        \end{align}
        \noindent recasting Equation (\ref{eq:received_echo_2}) as 
        \begin{align}
            \boldsymbol{Y} = \alpha \boldsymbol{c}(\tau)\left(\boldsymbol{d}^{\text{T}}(\nu) \odot \boldsymbol{g}^{\text{T}}(\boldsymbol{\theta}) \right) + \boldsymbol{Z} \in \mathbb{C}^{N_{c} \times Q L}, \label{eq:received_echo_3}
        \end{align}
        \noindent and by exploiting the repetitive structure of the \ac{IRS} phase-shift matrix in Equation (\ref{eq:IRS_profile_2}), it is possible to reformulate $\boldsymbol{g}(\boldsymbol{\theta})$ as
        \begin{align}
            \boldsymbol{g}^{\text{T}}(\boldsymbol{\theta}) = \underbrace{\left(\boldsymbol{b}^{\text{T}}(\boldsymbol{\theta}) [\boldsymbol{w}_{1} \cdots \boldsymbol{w}_{L}] \right)^{2}}_{\boldsymbol{g}^{\text{T}}_{L}(\boldsymbol{\theta}) \in \mathbb{C}^{1 \times L}} \otimes \boldsymbol{1}^{\text{T}}_{Q}, \label{eq:g_vector}
        \end{align}
        \noindent and plugging Equation (\ref{eq:g_vector}) into Equation (\ref{eq:received_echo_3}) leads to
        \begin{align}
            \boldsymbol{Y} = \alpha \boldsymbol{c}(\tau)\left[\boldsymbol{d}^{\text{T}}(\nu) \odot \left( \boldsymbol{g}^{\text{T}}_{L}(\boldsymbol{\theta}) \otimes \boldsymbol{1}^{\text{T}}_{Q} \right)\right] + \boldsymbol{Z}. 
            \label{eq:received_echo_4}
        \end{align}
        By defining $\boldsymbol{d}_{Q}(\nu) \in \mathbb{C}^{Q \times 1}$ and $\boldsymbol{d}_{L}(\nu) \in \mathbb{C}^{L \times 1}$ as 
        \begin{align}
            \boldsymbol{d}_{Q}(\nu) &= [1, e^{j 2 \pi \nu T_{s}}, \cdots, e^{j 2 \pi \nu (Q - 1) T_{s}}] \in \mathbb{C}^{Q \times 1}, \\
            \boldsymbol{d}_{L}(\nu) &= [1, e^{j 2 \pi \nu Q T_{s}}, \cdots, e^{j 2 \pi \nu Q (L - 1) T_{s}}] \in \mathbb{C}^{L \times 1},
        \end{align}
        \noindent allows us to express the time-domain steering vector as a function of the Doppler as $\boldsymbol{d}(v) = \boldsymbol{d}_{L}(\nu) \otimes \boldsymbol{d}_{Q}(\nu)$. Plugging this expression into Equation (\ref{eq:received_echo_4}) yields
        \begin{align}
            \notag \boldsymbol{Y} &\hspace{-0.05cm}=\hspace{-0.05cm} \alpha \boldsymbol{c}(\tau)\left[\left(\boldsymbol{d}^{\text{T}}_{L}(\nu) \otimes \boldsymbol{d}^{\text{T}}_{Q}(\nu)\right) \odot \left( \boldsymbol{g}^{\text{T}}_{L}(\boldsymbol{\theta}) \otimes \boldsymbol{1}^{\text{T}}_{Q} \right)\right] \hspace{-0.05cm}+\hspace{-0.05cm} \boldsymbol{Z}, 
        \end{align}
        and taking its transpose and applying property $\left(\boldsymbol{a} \otimes \boldsymbol{c}\right) \odot \left(\boldsymbol{b} \otimes \boldsymbol{d}\right) = \left(\boldsymbol{a} \odot \boldsymbol{b}\right) \otimes \left(\boldsymbol{c} \odot \boldsymbol{d}\right)$, leads to the following expression
        \begin{align}
            \notag \boldsymbol{Y} &\hspace{-0.05cm}=\hspace{-0.05cm} \alpha \boldsymbol{c}(\tau)\left[\left( \boldsymbol{g}_{L}(\boldsymbol{\theta}) \otimes \boldsymbol{1}_{Q} \right) \odot \left(\boldsymbol{d}_{L}(\nu) \otimes \boldsymbol{d}_{Q}(\nu)\right)\right]^{\text{T}} \hspace{-0.05cm}+\hspace{-0.05cm} \boldsymbol{Z}, \\
            \notag &\hspace{-0.05cm}=\hspace{-0.05cm} \alpha \boldsymbol{c}(\tau)\left[\left(\boldsymbol{g}_{L}(\boldsymbol{\theta}) \odot \boldsymbol{d}_{L}(\nu)\right) \otimes \left(\boldsymbol{1}_{Q} \odot \boldsymbol{d}_{Q}(\nu)\right)\right]^{\text{T}} \hspace{-0.05cm}+\hspace{-0.05cm} \boldsymbol{Z}, \\
            &\hspace{-0.05cm}=\hspace{-0.05cm} \alpha \boldsymbol{c}(\tau)\left[\left( \boldsymbol{g}_{L}(\boldsymbol{\theta}) \odot \boldsymbol{d}_{L}(\nu) \right) \otimes \boldsymbol{d}_{Q}(\nu)\right]^{\text{T}} \hspace{-0.05cm}+\hspace{-0.05cm} \boldsymbol{Z}. \label{eq:received_echo_5}
        \end{align}
        Defining $\boldsymbol{\beta}(\boldsymbol{\theta}, \nu) = \alpha\left(\boldsymbol{g}_{L}(\boldsymbol{\theta}) \odot \boldsymbol{d}_{L}(\nu) \right) \in \mathbb{C}^{L \times 1}$ and plugging it into Equation (\ref{eq:received_echo_5}) yields 
        \begin{align}
            \boldsymbol{Y} &= \boldsymbol{c}(\tau)\left(\boldsymbol{\beta}(\boldsymbol{\theta}, \nu) \otimes \boldsymbol{d}_{Q}(\nu) \right)^{\text{T}} + \boldsymbol{Z}, 
        \end{align}
         which can be viewed as the first-mode unfolding of the following rank-one third-order tensor $\ten{Y}$ 
        \begin{align}
            \ten{Y} = \boldsymbol{c}(\tau) \circ \boldsymbol{d}_{Q}(\nu) \circ \boldsymbol{\beta}(\boldsymbol{\theta}, \nu) + \ten{Z} \in \mathbb{C}^{N_{c} \times Q \times L} \label{eq:tensor_problem_1}
        \end{align}
        \noindent with its matrix unfoldings being written as \cite{comon2009tensor}
        \begin{align}
            \nmode{Y}{1} &= \boldsymbol{c}(\tau)\left(\boldsymbol{\beta}(\boldsymbol{\theta}, \nu) \otimes \boldsymbol{d}_{Q}(\nu) \right)^{\text{T}} \in \mathbb{C}^{N_{c} \times Q L}, \\
            \nmode{Y}{2} &= \boldsymbol{d}_{Q}(\nu) \left(\boldsymbol{\beta}(\boldsymbol{\theta}, \nu) \otimes \boldsymbol{c}(\tau) \right)^{\text{T}} \in \mathbb{C}^{Q \times N_{c} L}, \\
            \nmode{Y}{3} &= \boldsymbol{\beta}(\boldsymbol{\theta}, \nu) \left( \boldsymbol{d}_{Q}(\nu) \otimes \boldsymbol{c}(\tau) \right)^{\text{T}} \in \mathbb{C}^{L \times N_{c} Q}.
        \end{align}
        
\begin{figure}[!t]
            \centering
            \includegraphics[width=0.80\columnwidth]{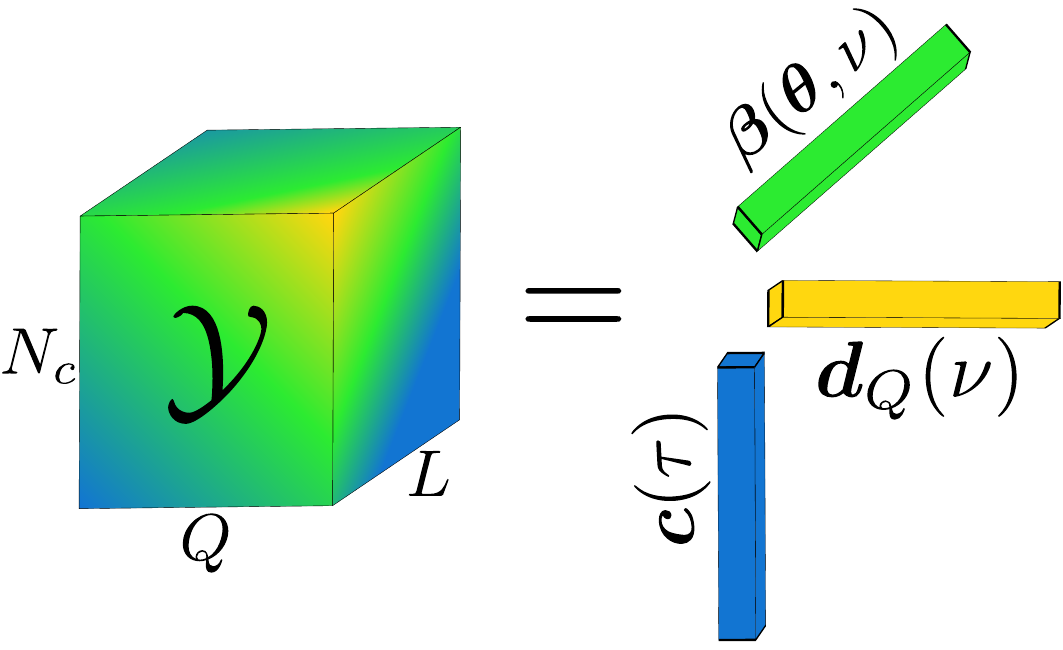}
            \caption{Noiseless received sensing signal tensor.}
            \label{fig:02}
        \end{figure}

        An illustration of the received sensing signal tensor (in the noiseless case) is shown in Figure \ref{fig:02}.
    \subsection{HOSVD-based Estimation}
    \indent Estimating the factors of the tensor in Equation (\ref{eq:tensor_problem_1}) requires us to find a multi-linear approximation of $\ten{Y}$. This can be achieved by employing the state-of-the-art truncated \ac{HOSVD} algorithm \cite{benicio2023tensor, comon2009tensor, de2016overview}. Let us define the following tensor minimization problem 
    \begin{equation}
        \underset{\boldsymbol{c}(\tau), \boldsymbol{d}_{Q}(\nu), \boldsymbol{\beta}(\boldsymbol{\theta}, \nu)} {\text{arg min}} \left|\left| \begin{split} \ten{Y} - \boldsymbol{c}(\tau) \circ \boldsymbol{d}_{Q}(\nu) \circ \boldsymbol{\beta}(\boldsymbol{\theta}, \nu) \end{split} \right|\right|^{2}_{\text{F}}. \label{eq:tensor_fit_1}
    \end{equation}     
    \noindent The optimization problem in Equation (\ref{eq:tensor_fit_1}) is solved by means of the \ac{HOSVD}, by computing multiple \acp{SVD} for each unfolding of $\ten{Y}$, as follows
    \begin{align}
        \nmode{Y}{1} &= \boldsymbol{U}^{(1)} \boldsymbol{\Sigma}^{(1)} \boldsymbol{V}^{(1)\text{H}}, \label{eq:svd1} \\
        \nmode{Y}{2} &= \boldsymbol{U}^{(2)} \boldsymbol{\Sigma}^{(2)} \boldsymbol{V}^{(2)\text{H}}, \label{eq:svd2}\\
        \nmode{Y}{3} &= \boldsymbol{U}^{(3)} \boldsymbol{\Sigma}^{(3)} \boldsymbol{V}^{(3)\text{H}}. \label{eq:svd3}
    \end{align}
    and the estimates of $\boldsymbol{c}(\tau)$, $\boldsymbol{d}_{Q}(\nu)$, and $\boldsymbol{\beta}(\boldsymbol{\theta}, \nu)$ are built from the dominant left singular vectors of $\nmode{Y}{1}$, $\nmode{Y}{2}$, and $\nmode{Y}{3}$, respectively, i.e.,
    \begin{align}
        \hat{\boldsymbol{c}}(\tau) = \boldsymbol{U}^{(1)}_{.1}, \,\, \hat{\boldsymbol{d}}_{Q}(\nu) = \boldsymbol{U}^{(2)}_{.1}, \,\, \boldsymbol{\beta}(\boldsymbol{\theta}, \nu) = \boldsymbol{U}^{(3)}_{.1}. \label{eq:factors}
    \end{align}
    \indent After the \ac{HOSVD} stage, we employ a one-dimensional peak search from $\hat{\boldsymbol{c}}(\tau)$ to estimate the delay $\tau$, a one-dimensional peak search from $\hat{\boldsymbol{d}}_{Q}(\nu)$ to estimate the Doppler $\nu$, and finally a peak search from $\hat{\boldsymbol{\beta}}(\boldsymbol{\theta}, \nu)$ to estimate the angular information between the \ac{IRS} and the target, $\boldsymbol{\theta}$. Since the received echo signal follows a rank-one tensor model, the \ac{HOSVD} leads to unique estimates up to a trivial scaling factor affecting each estimated vector. Note that the peak search problem is insensitive to such a scaling factor. Starting by $\tau$ and $\nu$, we solve the following peak search problems to obtain an estimation of delay and Doppler
    \begin{align}
         \hat{\tau} = \underset{\tau} {\text{arg max}} \frac{\left|\boldsymbol{c}^{\text{H}}(\tau) \hat{\boldsymbol{c}}(\tau)\right|^{2}}{\left|\hat{\boldsymbol{c}}(\tau)\right|^{2}}, \,\,
         \hat{\nu} = \underset{\nu}{\text{arg max}} \frac{\left|\boldsymbol{d}^{\text{H}}_{Q}(\nu) \hat{\boldsymbol{d}}_{Q}(\nu) \right|^{2}}{\left|\hat{\boldsymbol{d}}_{Q}(\nu)\right|^{2}}. \label{eq:delay_Doppler_search}
    \end{align}
    \indent Before solving the last peak search problem that estimates the angular information, we exploit knowledge of the Doppler estimated in Equation (\ref{eq:delay_Doppler_search}) to simplify the three-dimensional search in $\boldsymbol{\beta}(\boldsymbol{\theta}, \nu)$ to a two-dimensional search in $\boldsymbol{g}_{L}(\boldsymbol{\theta})$. To obtain an estimation of $\boldsymbol{g}_{L}(\boldsymbol{\theta})$ from $\hat{\boldsymbol{\beta}}(\boldsymbol{\theta}, \nu)$ we rewrite 
    \begin{align}
        \hat{\boldsymbol{g}_{L}(\boldsymbol{\theta})} = \hat{\boldsymbol{\beta}}(\boldsymbol{\theta}, \nu) \odot (\boldsymbol{d}_{L}(\nu))^{-1}, \label{eq:gL_estimation}
    \end{align} 
    and with the knowledge of the delay estimated in Equation (\ref{eq:delay_Doppler_search}), we formulate the following cost function \cite{kemal2024ris}
    \begin{align}
         \hat{\boldsymbol{\theta}} \hspace{-0.1cm}=\hspace{-0.1cm} \underset{\boldsymbol{\theta}} {\text{arg max}} \frac{\left|\boldsymbol{c}^{\text{H}}(\hat{\tau}) \hspace{-0.1cm}\left[\ten{Y}\right]_{(1)}\hspace{-0.1cm} \left[\left(\hat{\boldsymbol{g}}_{L}(\boldsymbol{\theta}) \hspace{-0.05cm}\odot\hspace{-0.05cm} \boldsymbol{d}_{L}(\hat{\nu})\right) \hspace{-0.05cm}\otimes\hspace{-0.05cm} \boldsymbol{d}_{Q}(\hat{\nu})\right]^{*}\right|^{2}}{\left|\hat{\boldsymbol{g}}_{L}(\boldsymbol{\theta})\right|^{2}}. \label{eq:angle_search}
    \end{align}
    \subsection{Complexity Analysis}
        In Table \ref{tab:complexity}, we derive the approximated computational complexity in terms of $\mathcal{O}(\cdot)$ for the selected baseline algorithm from \cite{kemal2024ris}, and our proposed solution summarized in Alg. \ref{alg:01}. The baseline proposed by the authors of \cite{kemal2024ris} solves two peak search problems, one to jointly estimate $\tau$ and $\nu$ and the other to estimate the azimuth and elevation components of $\boldsymbol{\theta}$. Regarding our proposed algorithm, the \ac{HOSVD}-based parameter estimation of Equation (\ref{eq:tensor_fit_1}) computes three rank-one \acp{SVD} followed by three peak search problems derived in Equation (\ref{eq:delay_Doppler_search}).
        We consider that $\mathcal{O}(N_{1}N_{2})$ is the complexity associated with the rank-one approximation of a matrix $\mathbf{A} \in \mathbb{C}^{N_{1} \times N_{2}}$ by the \ac{SVD} \cite{benicio2023tensor_wcl}. Also, the complexity associated with the computation of the inner product between two vectors $\boldsymbol{a} \in \mathbb{C}^{N_{1} \times 1}$ and $\boldsymbol{b} \in \mathbb{C}^{N_{2} \times 1}$ and the product between a vector $\boldsymbol{a} \in \mathbb{C}^{N_{1} \times 1}$ and a matrix $\boldsymbol{A} \in \mathbb{C}^{N_{1} \times N_{2}}$ is given by $(2N_{1} - 1)$ and $(2 N_{1} N_{2} - N_{2})$, respectively \cite{hunger2005floating}. We use these definitions to compute the approximate number of flops linked to each iteration needed to solve the peak search problems of both the baseline \cite{kemal2024ris} and the proposed Alg. \ref{alg:01} considering that $R_{\tau}$, $R_{\nu}$, and $R_{\boldsymbol{\theta}}$ are the number of points of each grid defined by the peak search problems of $\tau$, $\nu$, and $\boldsymbol{\theta}$, respectively. The baseline \cite{kemal2024ris} complexity consists of solving the two-dimensional peak search problems that estimate $\tau$ and $\nu$, and the $\boldsymbol{\theta}$. Meanwhile the complexity of the proposed solution in Alg. \ref{alg:01} consists of solving the tensor estimation in Equation \ref{eq:tensor_fit_1} employing the \ac{HOSVD} followed by two one-dimensional peak search problems, Equation (\ref{eq:delay_Doppler_search}), and a single two-dimensional peak search problem in Equation (\ref{eq:angle_search}). From Table \ref{tab:complexity}, we can observe that the proposed Alg. \ref{alg:01} has a lower complexity when compared to the baseline \cite{kemal2024ris} since it can solve the peak search problem of $\tau$ and $\nu$ in parallel.
        \begin{algorithm}[!t]
            \small
            \caption{\ac{HOSVD}-based parameter estimation}
            \label{alg:01}
            \begin{algorithmic}[1]
                \Require{Tensor $\ten{Y}$}
                \State{Define the \ac{SVD}s of $\nmode{Y}{1}$, $\nmode{Y}{2}$, and $\nmode{Y}{3}$ with (\ref{eq:svd1})-(\ref{eq:svd3}).}
                \State{Compute the estimation of $\boldsymbol{c}(\tau)$, $\boldsymbol{d}_{Q}(\nu)$, and $\boldsymbol{\beta}(\boldsymbol{\theta}, \nu)$ with (\ref{eq:factors}).}
                \State{Compute the estimation of $\tau$ and $\nu$ with (\ref{eq:delay_Doppler_search})}
                \State{Compute the estimation of $\boldsymbol{g}_{L}(\boldsymbol{\theta})$ with (\ref{eq:gL_estimation})}
                \State{Compute the estimation of $\boldsymbol{\theta}$ with (\ref{eq:angle_search})}
                \State \textbf{return} $\hat{\tau}$, $\hat{\nu}$, and $\hat{\boldsymbol{\theta}}$
            \end{algorithmic}
        \end{algorithm}
        \begin{table*}[!t]
            \centering
            \caption{Computational complexity of the baseline and the proposed algorithms}
            \label{tab:complexity}
            \resizebox{0.85\textwidth}{!}{
            \begin{tabular}{|c|c|}
            \hline
            \textbf{Algorithm} & \textbf{Computational Complexity} \\ \hline
            Baseline \cite{kemal2024ris} & $\mathcal{O}(\underbrace{R_{\tau} R_{\nu} L (2 N_{c} Q + N_{c} - 1)}_{\tau \text{ and } \nu \text{ estimation}} + \underbrace{R_{\boldsymbol{\theta}} (2 N_{c} Q + N_{c} - 1))}_{\boldsymbol{\theta} \text{ estimation}}$ \\ \hline
            Proposed &  $\mathcal{O}(\underbrace{3 N_{c} Q L}_{\text{HOSVD in } \ref{eq:tensor_fit_1}} + \underbrace{R_{\tau} (2N_{c} - 1)}_{\text{Equation } \ref{eq:delay_Doppler_search}} + \underbrace{R_{\nu} (2 Q L - 1)}_{\text{Equation } \ref{eq:delay_Doppler_search}} + \underbrace{R_{\boldsymbol{\theta}} (2 N_{c} Q + N_{c} - 1))}_{\text{Equation } \ref{eq:angle_search}}$ \\ \hline
            \end{tabular}
            }
        \end{table*}
\section{Simulation Results}
\begin{figure}[!t]
        \centering
        \includegraphics[width = 0.75\linewidth]{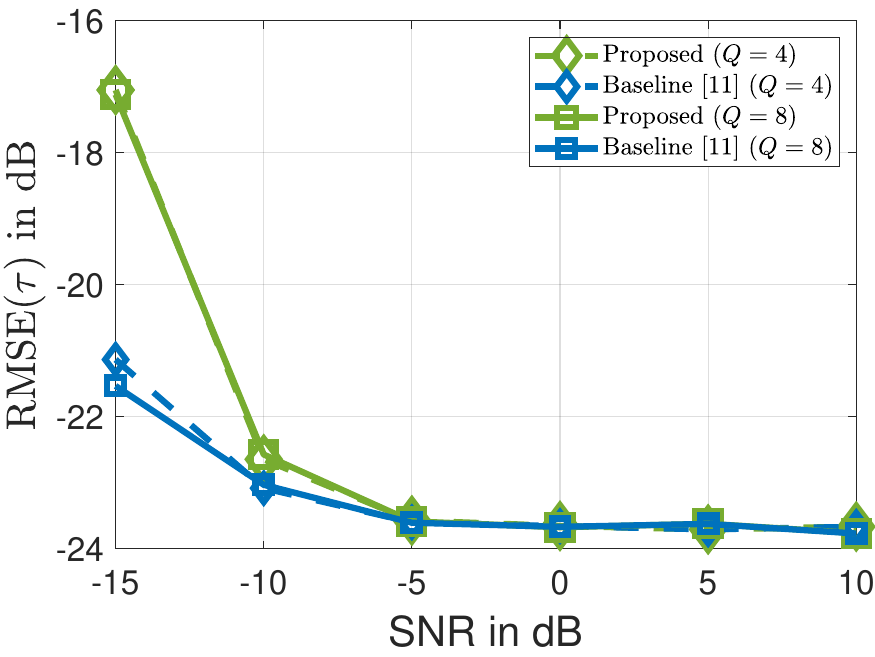}
        \caption{Delay estimation performance of the proposed solution and the baseline algorithm in \cite{kemal2024ris}.}
        \label{fig:04}
    \end{figure}

    \indent We evaluate the performance of the proposed tensor-based algorithm by comparing it with the parameter estimation method on \cite{kemal2024ris} in terms of both \ac{RMSE} and computational complexity. We design the \ac{IRS} phase-shift matrix $\boldsymbol{W}$ as a truncated \ac{DFT} of size $N$ with $L$ columns. The \ac{IRS} elevation and azimuth angles of arrival, $\boldsymbol{\phi}$, and departure, $\boldsymbol{\theta}$, are randomly generated from a uniform distribution between $[0, \pi/2]$. The parameter estimation accuracy is evaluated in terms of the \ac{RMSE} given by $\text{\ac{RMSE}}(x) = \sqrt{\mathbb{E}\left[\left|x^{(m)} - \hat{x}^{(m)}\right|^{2}/\left|x^{(m)}\right|^{2}\right]}$ with $x$ being one of the estimated parameters at the $m$th experiment, with $V = 5 \times 10^{3}$ being the number of channel realizations. Also, the \ac{SNR} is defined as $\text{SNR} = ||\ten{Y}||^{2}_{\text{F}}\sigma^{2}_{\ten{Z}}/||\ten{Z}||^{2}_{\text{F}}$ with $\sigma^{2}_{\ten{Z}}$ being the noise variance. Table \ref{tab:parameters} summarizes the remaining simulation parameters. In Fig. \ref{fig:04}, we evaluate the \ac{RMSE} performance of delay estimation for the baseline algorithm in \cite{kemal2024ris} and the proposed Algorithm \ref{alg:01} for two different scenarios, $Q = 4$ and $Q = 8$. The baseline solution performs slightly better in the very low \ac{SNR} range. Note that the performance of both methods is the same at \ac{SNR} levels higher than $-5$ dB. It can also be seen that the delay estimation performance of both methods is not influenced by the block size $Q$.  

\begin{figure}[!t]
        \centering
        \includegraphics[width = 0.75\linewidth]{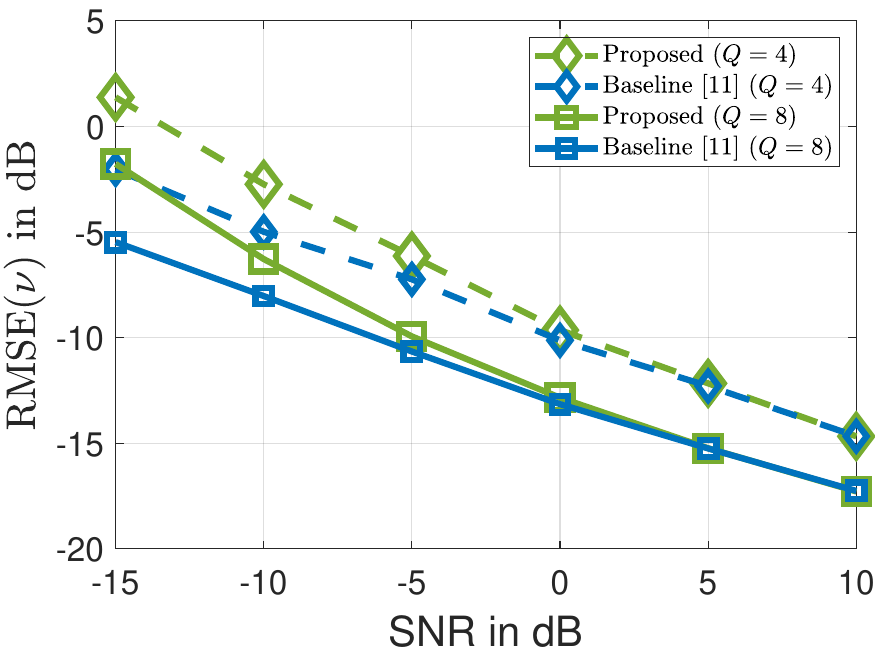}
        \caption{Doppler estimation performance of the proposed solution and the baseline algorithm in \cite{kemal2024ris}.}
        \label{fig:05}
    \end{figure}

    \indent Similarly, in Fig. \ref{fig:05} we evaluate the \ac{RMSE} performance of Doppler estimation for the baseline algorithm in \cite{kemal2024ris} and the proposed Algorithm \ref{alg:01} for the two same scenarios, $Q= 4$ and $Q= 8$. The proposed solution performs similarly to the baseline for most of the considered SNR range while exhibiting some performance degradation for very low SNR levels. Indeed, the proposed method, although much simpler than the one of \cite{kemal2024ris}, is a bit more sensitive to very high noise. We can also note that increasing the block size $Q$ (or, equivalently, the repetition factor of the IRS phase shifts) improves performance for both methods.

\begin{figure}[!t]
        \centering
        \includegraphics[width = 0.75\linewidth]{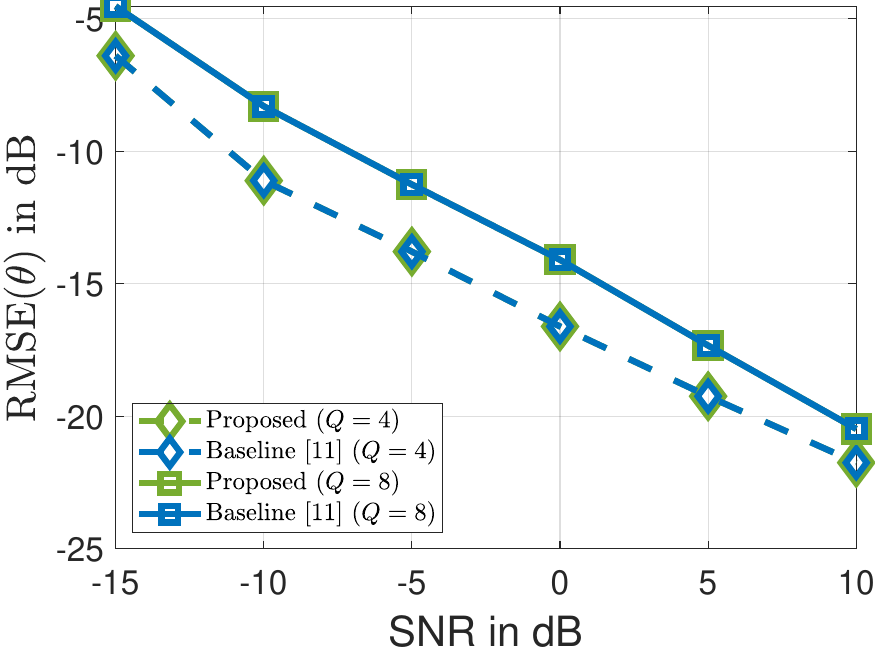}
        \caption{Angle estimation performance of the proposed solution and the baseline algorithm in \cite{kemal2024ris}.}
        \label{fig:06}
    \end{figure}

    \indent In Fig. \ref{fig:06} we observe the angular estimation \ac{RMSE} performance for the baseline algorithm in \cite{kemal2024ris} and the proposed tensor-based sensing algorithm for the scenarios $Q = 4$ and $Q = 8$. In this case, both the baseline in \cite{kemal2024ris} and the proposed solution have the same performance for both scenarios, while the increase in $Q$ harms the estimation for both methods. Note that both the proposed and the baseline methods extract the angular estimates using a maximum likelihood criterion (see equation \ref{eq:angle_search}), which explains their similar performances. Note, however, that the proposed method has an overall lower complexity due to the decoupling of the delay and Doppler estimates achieved after the HOSVD stage.
    
\begin{figure}[!t]
        \centering
        \includegraphics[width = 0.75\columnwidth]{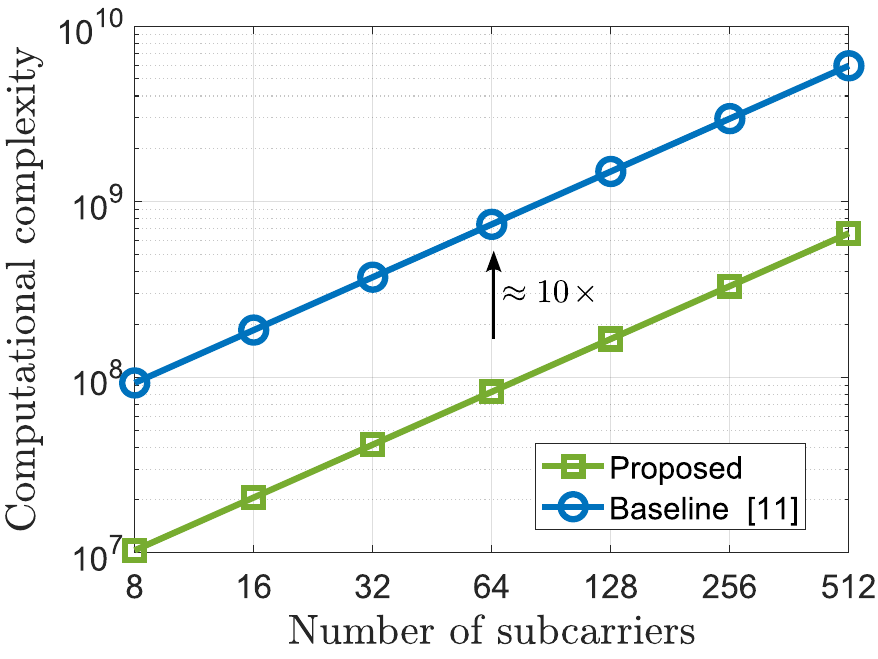}
        \caption{Computational complexity comparison as a function of the number of subcarriers.}
        \label{fig:07}
    \end{figure}

\begin{figure}[!t]
        \centering
        \includegraphics[width = 0.75\columnwidth]{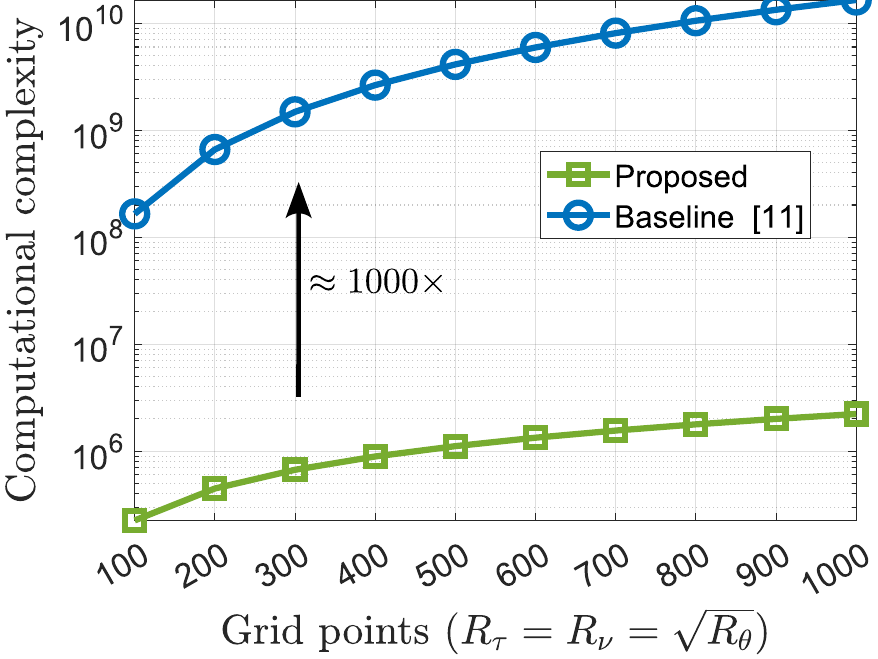}
        \caption{Computational complexity comparison as a function of the number of grid points.}
        \label{fig:08}
    \end{figure}

    \indent In Fig. \ref{fig:07}, we show the computational complexity curves for the proposed solution in Algorithm \ref{alg:01} and the baseline solution in \cite{kemal2024ris} as a function of the number of subcarriers. The other parameters are fixed according to Table \ref{tab:parameters}. The baseline method is approximately $10$ times more complex than the proposed tensor-based method, while the complexity increases linearly with the number of subcarriers. This behavior is explained by analyzing Table \ref{tab:complexity}. The complexity of the reference method is dependent on the product $R_{\tau} R_{\nu}$, which corresponds to the size of the grids defined by the peak search problems that estimate the delay and the Doppler since there is a coupling estimate of the delay and the Doppler in \cite{kemal2024ris}. However, Algorithm \ref{alg:01} can estimate the delay and the Doppler in parallel, which breaks the product $R_{\tau} R_{\nu}$ into a summation. Similarly, in Fig. \ref{fig:08}, we plot the complexity curve according to Table \ref{tab:complexity} as a function of the number of points in each grid that solves the peak search problems. These results indicate that the proposed tensor solution is approximately $10$ times less complex than the competing method \cite{kemal2024ris} for a varying number of subcarriers.  
    \begin{table}[!t]
        \centering
        \caption{Simulation parameters.}
        \label{tab:parameters}
        \resizebox{0.975\columnwidth}{!}{
        \begin{tabular}{|c|c|}
            \hline
            \ac{IRS} size & $N_{x} N_{y} = 4 \times 4 = 16$ \\ \hline
            \ac{IRS} spacing & $d_{x} = d_{y} = \lambda/2$ \\ \hline
            \ac{AoA} generation & $\{\phi_{\text{az}}, \phi_{\text{el}}\} \sim \text{U}(0, 90^{\circ})$ \\ \hline
            \ac{AoD} generation & $\{\theta_{\text{az}}, \theta_{\text{el}}\} \sim \text{U}(0, 90^{\circ})$ \\ \hline
            Wavelength & $1.07 \times 10^{-2}$ m \\ \hline
            Carrier frequency & $28$ GHz \\ \hline
            Subcarrier spacing $\Delta f$ & $120$ KHz \\ \hline
            Symbol duration & $1/\Delta f$ \\ \hline
            Channel realizations & $5000$ \\ \hline
            Number of symbols & $64$ \\ \hline
            Number of subcarriers & $16$ \\ \hline
            Distance \ac{BS} - \ac{IRS} & $10$ m \\ \hline
            Distance \ac{IRS} - target & $5$ m \\ \hline
            \Acf{RCS} & $2 \text{m}^{2}$ \\ \hline
            Grid size for (\ref{eq:delay_Doppler_search}) & $R_{\tau} = R_{\nu} = 100$ \\ \hline
            Grid size for (\ref{eq:angle_search}) & $R_{\boldsymbol{\theta}} = 10000$ \\ \hline
        \end{tabular}
        }
    \end{table}
\section{Conclusion}
    This paper focuses on the parameter estimation problem in a monostatic sensing system assisted by IRS. It proposes a tensor-based estimation algorithm, summarized in Alg. \ref{alg:01}, that estimates the delay, Doppler, and angles in an \ac{IRS}-assisted scenario. The proposed solution in Alg. \ref{alg:01} performs slightly worse for delay and Doppler estimation at medium \ac{SNR}, while at high and low \ac{SNR}, both algorithms show similar performance. Regarding the \ac{AoA} estimation, the baseline solution in \cite{kemal2024ris} and the proposed tensor-based method achieve the same performance in terms of \ac{RMSE}. However, the proposed solution substantially reduces the computational complexity and improves scalability: with increasing subcarriers, our solution is $10$ times less complex than the baseline algorithm. As a perspective, we will investigate the performance of the proposed tensor-based solution by considering the intrinsic self-interference of monostatic sensing scenarios.

\bibliographystyle{IEEEtran}
\bibliography{addons/bibliography}

\end{document}

%% file: addons/preamble.tex
\usepackage{algorithm,algpseudocode}
\usepackage{amstext,amsmath,amssymb,exscale,mathtools,bm,gensymb,cite}
\usepackage{psfrag}
\usepackage[none]{hyphenat}
\usepackage{tabularx}
\usepackage[shortlabels]{enumitem}
\usepackage{xcolor}
\usepackage{caption,subcaption,siunitx}
\usepackage[nolist,nohyperlinks]{acronym}
\usepackage{scalerel}
\usepackage{makecell}
\usepackage{graphicx} 
\usepackage{cite}
\usepackage{textcomp}

\def\BibTeX{{\rm B\kern-.05em{\sc i\kern-.025em b}\kern-.08em
    T\kern-.1667em\lower.7ex\hbox{E}\kern-.125emX}}

\usepackage{tikz,tikz-3dplot,tkz-tab}
\usetikzlibrary{angles, calc, positioning, quotes, intersections,fit,decorations.pathreplacing,decorations.markings}
\tdplotsetmaincoords{70}{120}

\tikzset{
    cross/.style={fill=white,path picture={\draw[black]
        (path picture bounding box.south east) -- (path picture bounding box.north west)
        (path picture bounding box.south west) -- (path picture bounding box.north east);}},
    dressed/.style={fill=white,postaction={pattern=north east lines}},
    momentum/.style={->,semithick,yshift=5pt,shorten >=5pt,shorten <=5pt},
    loop/.style 2 args={thick,decoration={markings,mark=at position {#1} with {\arrow{<},\node[anchor=\pgfdecoratedangle-90,font=\footnotesize] {};}},postaction={decorate}},
    label/.style={thin,gray,shorten <=-1.5ex}
}

\tikzset{
    cross/.pic = {
    \draw[rotate = 45] (-#1,0) -- (#1,0);
    \draw[rotate = 45] (0,-#1) -- (0, #1);
    }
}

%% file: manuscript_arxiv.bbl
\begin{thebibliography}{10}
\providecommand{\url}[1]{#1}
\csname url@samestyle\endcsname
\providecommand{\newblock}{\relax}
\providecommand{\bibinfo}[2]{#2}
\providecommand{\BIBentrySTDinterwordspacing}{\spaceskip=0pt\relax}
\providecommand{\BIBentryALTinterwordstretchfactor}{4}
\providecommand{\BIBentryALTinterwordspacing}{\spaceskip=\fontdimen2\font plus
\BIBentryALTinterwordstretchfactor\fontdimen3\font minus
  \fontdimen4\font\relax}
\providecommand{\BIBforeignlanguage}[2]{{%
\expandafter\ifx\csname l@#1\endcsname\relax
\typeout{** WARNING: IEEEtran.bst: No hyphenation pattern has been}%
\typeout{** loaded for the language `#1'. Using the pattern for}%
\typeout{** the default language instead.}%
\else
\language=\csname l@#1\endcsname
\fi
#2}}
\providecommand{\BIBdecl}{\relax}
\BIBdecl

\bibitem{zheng2022survey}
B.~Zheng \emph{et~al.}, ``A survey on channel estimation and practical passive
  beamforming design for intelligent reflecting surface aided wireless
  communications,'' \emph{IEEE Commun. Surv. Tutor.}, vol.~24, no.~2, pp.
  1035--1071, 2022.

\bibitem{pan2022overview}
C.~Pan \emph{et~al.}, ``An overview of signal processing techniques for
  {RIS}/{IRS}-aided wireless systems,'' \emph{IEEE J. Sel. Topics Signal
  Processing}, vol.~16, no.~5, pp. 883--917, 2022.

\bibitem{astrom2024ris}
M.~{\AA}str{\"o}m \emph{et~al.}, ``Ris in cellular networks--challenges and
  issues,'' \emph{arXiv preprint arXiv:2404.04753}, 2024.

\bibitem{wu2021intelligent}
Q.~Wu \emph{et~al.}, ``Intelligent reflecting surface-aided wireless
  communications: A tutorial,'' \emph{IEEE transactions on communications},
  vol.~69, no.~5, pp. 3313--3351, 2021.

\bibitem{liu2022integrated}
F.~Liu \emph{et~al.}, ``Integrated sensing and communications: Toward
  dual-functional wireless networks for {6G} and beyond,'' \emph{IEEE Journal
  on Selected areas in Comms.}, vol.~40, no.~6, pp. 1728--1767, 2022.

\bibitem{chepuri2023integrated}
S.~P. Chepuri \emph{et~al.}, ``Integrated sensing and communications with
  reconfigurable intelligent surfaces: From signal modeling to processing,''
  \emph{IEEE Signal Processing Magazine}, vol.~40, no.~6, pp. 41--62, 2023.

\bibitem{zhang2022metaradar}
H.~Zhang \emph{et~al.}, ``{MetaRadar}: Multi-target detection for
  reconfigurable intelligent surface aided radar systems,'' \emph{IEEE
  Transactions on Wireless Communications}, vol.~21, no.~9, pp. 6994--7010,
  2022.

\bibitem{rihan2022spatial}
M.~Rihan \emph{et~al.}, ``Spatial diversity in radar detection via active
  reconfigurable intelligent surfaces,'' \emph{IEEE Signal Processing Letters},
  vol.~29, pp. 1242--1246, 2022.

\bibitem{song2023intelligent}
X.~Song \emph{et~al.}, ``Intelligent reflecting surface enabled sensing:
  Cram{\'e}r-{Rao} bound optimization,'' \emph{IEEE Transactions on Signal
  Processing}, 2023.

\bibitem{shao2024target}
X.~Shao \emph{et~al.}, ``Target-mounted intelligent reflecting surface for
  secure wireless sensing,'' \emph{IEEE Transactions on Wireless
  Communications}, 2024.

\bibitem{kemal2024ris}
M.~Kemal~Ercan \emph{et~al.}, ``{RIS}-aided {NLoS} monostatic sensing under
  mobility and angle-doppler coupling,'' \emph{IEEE Wireless Communications and
  Networking Conference 2024}, 2024.

\bibitem{Asim_2021}
Fazal-E-Asim \emph{et~al.}, ``Two-dimensional channel parameter estimation for
  millimeter-wave systems using butler matrices,'' \emph{IEEE Transactions on
  Wireless Communications}, vol.~20, no.~4, pp. 2670--2684, 2021.

\bibitem{comon2009tensor}
P.~Comon \emph{et~al.}, ``Tensor decompositions, alternating least squares and
  other tales,'' \emph{Journal of Chemometrics: A Journal of the Chemometrics
  Society}, vol.~23, no. 7-8, pp. 393--405, 2009.

\bibitem{benicio2023tensor}
K.~Ben{\'\i}cio \emph{et~al.}, ``Tensor-based modeling/estimation of static
  channels in {IRS}-assisted {MIMO} systems,'' \emph{XLI Brazilian Symposium on
  Telecommunications and Signal Processing - SBrT 2023}, 2023.

\bibitem{de2016overview}
A.~L.~F. de~Almeida \emph{et~al.}, ``{Overview of tensor decompositions with
  applications to communications},'' in \emph{{Signals and Images: Advances and
  Results in Speech, Estimation, Compression, Recognition, Filtering, and
  Processing}}, R.~Coelho \emph{et~al.}, Eds.\hskip 1em plus 0.5em minus
  0.4em\relax {CRC-Press}, 1 2016, no. Chapter 12, pp. 325--356.

\bibitem{benicio2023tensor_wcl}
K.~B. Ben{\'\i}cio \emph{et~al.}, ``Tensor-based channel estimation and
  data-aided tracking in {IRS}-assisted {MIMO} systems,'' \emph{IEEE Wireless
  Comms. Letters}, 2023.

\bibitem{hunger2005floating}
R.~Hunger, \emph{Floating point operations in matrix-vector calculus}.\hskip
  1em plus 0.5em minus 0.4em\relax Munich University of Technology, Inst. for
  Circuit Theory, 2005, vol. 2019.

\end{thebibliography}
